\documentstyle[12pt]{article}

\catcode`\@=11
\long\def\@makefntext#1{
\protect\noindent \hbox to 3.2pt {\hskip-.9pt
$^{{\ninerm\@thefnmark}}$\hfil}#1\hfill}		

\def\@makefnmark{\hbox to 0pt{$^{\@thefnmark}$\hss}}  

\def\ps@myheadings{\let\@mkboth\@gobbletwo
\def\@oddhead{\hbox{}
\rightmark\hfil\ninerm\thepage}
\def\@oddfoot{}\def\@evenhead{\ninerm\thepage\hfil
\leftmark\hbox{}}\def\@evenfoot{}
\def\sectionmark##1{}\def\subsectionmark##1{}}

\newcounter{sectionc}\newcounter{subsectionc}\newcounter{subsubsectionc}
\renewcommand{\section}[1] {\vspace*{0.6cm}\addtocounter{sectionc}{1}
\setcounter{subsectionc}{0}\setcounter{subsubsectionc}{0}\noindent
	{\normalsize\bf\thesectionc\hspace{0.5em} #1}\par\vspace*{0.4cm}}
\renewcommand{\subsection}[1] {\vspace*{0.6cm}\addtocounter{subsectionc}{1}
	\setcounter{subsubsectionc}{0}\noindent
	{\normalsize\it\thesectionc.\thesubsectionc. #1}\par\vspace*{0.4cm}}
\renewcommand{\subsubsection}[1]
{\vspace*{0.6cm}\addtocounter{subsubsectionc}{1}
	\noindent {\normalsize\rm\thesectionc.\thesubsectionc.\thesubsubsectionc.
	#1}\par\vspace*{0.4cm}}

\newcounter{appendixc}
\newcounter{subappendixc}[appendixc]
\newcounter{subsubappendixc}[subappendixc]

\renewcommand{\appendix}[1] {\vspace*{0.6cm}
        \refstepcounter{appendixc}
        \setcounter{figure}{0}
        \setcounter{table}{0}
        \setcounter{equation}{0}
        \renewcommand{\thefigure}{\Alph{appendixc}.\arabic{figure}}
        \renewcommand{\thetable}{\Alph{appendixc}.\arabic{table}}
        \renewcommand{\theappendixc}{\Alph{appendixc}}
        \renewcommand{\theequation}{\Alph{appendixc}.\arabic{equation}}
        \noindent{\bf Appendix \theappendixc #1}\par\vspace*{0.4cm}}

\def\abstracts#1{{

\centering{\begin{minipage}{15.2truecm}\footnotesize\baselineskip=12pt\noindent
	\parindent=0pt #1
	\end{minipage}}\par}}

\renewenvironment{thebibliography}[1]
	{\begin{list}{\arabic{enumi}.}
	{\usecounter{enumi}\setlength{\parsep}{0pt}
\setlength{\leftmargin 0.8cm}{\rightmargin 0pt}
	 \setlength{\itemsep}{0pt} \settowidth
	{\labelwidth}{#1.}\sloppy}}{\end{list}}

\newcounter{itemlistc}
\newcounter{romanlistc}
\newcounter{alphlistc}
\newcounter{arabiclistc}

\newcommand{\fcaption}[1]{
        \refstepcounter{figure}
        \setbox\@tempboxa = \hbox{\footnotesize Fig.~\thefigure. #1}
        \ifdim \wd\@tempboxa > 6in
           {\begin{center}
        \parbox{6in}{\footnotesize\baselineskip=12pt Fig.~\thefigure. #1}
            \end{center}}
        \else
             {\begin{center}
             {\footnotesize Fig.~\thefigure. #1}
              \end{center}}
        \fi}

\newcommand{\tcaption}[1]{
        \refstepcounter{table}
        \setbox\@tempboxa = \hbox{\footnotesize Table~\thetable. #1}
        \ifdim \wd\@tempboxa > 6in
           {\begin{center}
        \parbox{6in}{\footnotesize\baselineskip=12pt Table~\thetable. #1}
            \end{center}}
        \else
             {\begin{center}
             {\footnotesize Table~\thetable. #1}
              \end{center}}
        \fi}

\def\@citex[#1]#2{\if@filesw\immediate\write\@auxout
	{\string\citation{#2}}\fi
\def\@citea{}\@cite{\@for\@citeb:=#2\do
	{\@citea\def\@citea{,}\@ifundefined
	{b@\@citeb}{{\bf ?}\@warning
	{Citation `\@citeb' on page \thepage \space undefined}}
	{\csname b@\@citeb\endcsname}}}{#1}}

\newif\if@cghi
\def\cite{\@cghitrue\@ifnextchar [{\@tempswatrue
	\@citex}{\@tempswafalse\@citex[]}}
\def\citelow{\@cghifalse\@ifnextchar [{\@tempswatrue
	\@citex}{\@tempswafalse\@citex[]}}
\def\@cite#1#2{{$\null^{#1}$\if@tempswa\typeout
	{IJCGA warning: optional citation argument
	ignored: `#2'} \fi}}

 1
 1
 1

\font\ninerm=cmr9

\newcommand{\beqn}{\begin{equation}}
\newcommand{\eeqn}{\end{equation}}
\newcommand{\barr}{\begin{eqnarray}}
\newcommand{\earr}{\end{eqnarray}}

\begin{document}
\centerline{\bf\large CHIRAL SYMMETRY RESTORATION AT FINITE TEMPERATURE}
\vspace{0.3cm}
\centerline{\bf\large IN THE DUAL GINZBURG-LANDAU THEORY}
\vspace{1.0cm}
\centerline{S. SASAKI, H. SUGANUMA and H. TOKI}
\vspace{0.1cm}
\centerline{\it Research Center for Nuclear Physics (RCNP)
, Osaka University, Osaka 567, JAPAN}
\vspace{0.6cm}
\abstracts{We study the relation between QCD-monopole condensation
and dynamical chiral-symmetry breaking in the dual Ginzburg-Landau Theory,
which realizes the color confinement due to the dual Meissner effect.
Solving the Schwinger-Dyson equation at finite temperature numerically,
we find the chiral symmetry is restored at high temperature.
The critical temperature of the chiral symmetry restoration is
strongly correlated with the string tension.}
\normalsize\baselineskip=14pt
\vspace {0.6cm}
In 1981, 'tHooft pointed out the appearance of magnetic monopoles in
the abelian gauge based on the topological argument,
and conjecture that the dual Meissner effect
would be brought if QCD-monopoles are condensed\cite{tHooft}.
The color-electric flux between quarks is squeezed like a string or a tube
because the color-electric fields are excluded in the QCD-vacuum.
The linear-confining potential is realized since the squeezed flux
has the uniform energy per unit length.
The recent lattice QCD simulation shows that QCD-monopole condensation plays a
crucial role on color confinement\cite{Latt94}.
In the dual Ginzburg-Landau (DGL) theory\cite{kanazawa}, QCD-monopole
condensation causes the strong and long range correlation
between quark and antiquark,
which produces the linear-confining potential with the string tension as
$\sigma \simeq {{\footnotesize {{\bf Q}^2 m_{B}^2} \over 8\pi}}
\ln(1+{\footnotesize {m_{\chi}^2 \over m_{B}^2}})$
through the dual Higgs mechanism\cite{SST}.

We have studied the chiral symmetry breaking in the QCD-monopole
condensed vacuum by using the Schwinger-Dyson (SD) equation for the
dynamical quark, and found that QCD-monopole condensation plays an essential
role
on the chiral symmetry breaking as well\cite{{SST},{SST2}}.
In this paper, we study the manifestation of chiral symmetry at finite
temperature in the QCD vacuum.

Our strategy is to use the gluon propagator in
the DGL theory as the full gluon propagator of QCD in the SD equation
in order to include the non-perturbative effect in infrared region.
Taking the rainbow approximation, we get the SD equation for
the dynamical quark propagator $S_q(p)$ in the chiral limit,
\beqn
S_q^{-1}(p)= i{p \kern -2mm /} +
\int {{d^4k} \over (2\pi)^4}
{\bf Q}^2 \gamma_\mu S_q(k) \gamma_\nu D_{\mu \nu}
(p-k) \; ; \;{\bf Q}^2 = {{N_c -1} \over {2 N_c}}\cdot e^2
\label{sde}
\eeqn
in the Euclidean metric.
The gluon propagator $D_{\mu \nu}(k)$ in the QCD-monopole condensed
vacuum is derived from the DGL Lagrangian,
\beqn
D_{\mu \nu}(k)=-{1 \over k^2} \left( {\delta_{\mu \nu}+(\alpha_e -1)
{{k_\mu k_\nu} \over k^2}} \right) - {1 \over k^2}{m^2_B \over {k^2 + m^2_B}}
\cdot
{{\epsilon_{\lambda \mu \alpha \beta} {\epsilon_{\lambda \nu \gamma \delta}}
n_\alpha n_\gamma k_\beta k_\delta} \over {{(n \cdot k)}^2 + a^2}} \;,
\label{GP}
\eeqn
where $\alpha_e$ is the gauge fixing parameter on the residual abelian gauge
symmetry.
The mass of the dual gluon, $m_B$, is generated through the dual Higgs
mechanism
when QCD-monopoles are condensed\cite{{kanazawa},{SST}}.
We introduce the infrared cutoff $a$ in this propagator (\ref{GP})
corresponding
to the dynamical quark-antiquark pair creation and/or the size of
hadrons\cite{SST2}.
We take the angular average on the direction of
the Dirac string $n_\mu$ in the SD equation,
\beqn
\left\langle {1 \over (n\cdot k)^2+a^2} \right\rangle_{\rm average}
 \equiv {1 \over 2\pi ^2}\int {d\Omega_n}
          {1 \over (n\cdot k)^2+a^2}
 = {2 \over a }\cdot{1 \over
 a + \sqrt {k^2+a^2}} \; .
\label{average}
\eeqn
Here, the dynamical quark is considered to move in various
directions inside hadrons, and hence the constituent quark mass would be
regarded as
the quark self-energy in the angle-averaged case\cite{SST2}.
Taking a simple form for the quark propagator
as $S_q^{-1}(p)=i{p \kern -2mm /}- M(p^2)$,
the SD equation for the quark self-energy $M(p^2)$ is obtained
by taking the trace of Eq.(\ref{sde}),
\barr
M(p^2)&=& \int {d^4k \over (2\pi)^4} {\bf Q}^2
{M(k^2) \over k^2+M^2(k^2)} \cr
&& \times \left[{2 \over {\tilde k}^2+m_B^2}+{1+\alpha_e \over {\tilde k}^2}
+{4 \over a }\cdot{1 \over
 a + \sqrt {{\tilde k}^2+a^2}} \left( { {m^2_B-a^2}
\over {{\tilde k^2}+m^2_B} }
+ {a^2 \over {\tilde k^2}}
\right) \right]
\label{sde2}
\earr
with ${\tilde k}_\mu \equiv p_\mu - k_\mu$. It is noted that the r.h.s.
of Eq.(\ref{sde2}) is always non-negative.

We solve Eq.(\ref{sde2}) numerically using the Higashijima-Miransky
approximation with the QCD scale parameter $\Lambda_{\rm QCD}$
fixed at $200{\rm MeV}$,
and get the quark self-energy $M(p^2)$
as the function of the Euclidean momentum squared in a unit of
$\Lambda_{\rm QCD}$ at various $m_B$ as shown in Fig.1.
One finds that QCD-monopole condensation provides a large
contribution to spontaneous chiral-symmetry breaking\cite{{SST},{SST2}},
because the dual gluon mass $m_B$ is proportional to the QCD-monopole
condensate\cite{{kanazawa},{SST}} and the quark self-energy increases with
$m_B$ at each $p^2$.
The parameters are set as
$e=5.5$, $m_B=0.5{\rm GeV}$ and $a=85{\rm MeV}$
, which reproduce the string tension ($\sqrt{\sigma} \simeq 0.45 {\rm GeV}$)
and the
cylindrical radius of the hadron flux tube ($R\sim m_B^{-1} \simeq 0.4 {\rm
fm}$),
where these parameters are almost set by fitting the confinement phenomena.
The quark condensate and the pion decay constant are well-reproduced as
$\langle \bar qq \rangle_{_{\rm RGI}} \simeq-(247{\rm MeV})^3$ and $f_\pi
\simeq 88{\rm MeV}$, respectively.

\vspace{0.3cm}
\newlength{\minitwocolumn}
\setlength{\minitwocolumn}{4.1in}
\begin{minipage}{\minitwocolumn}
\end{minipage}
\hspace{\columnsep}
\setlength{\minitwocolumn}{2.3in}
\begin{minipage}{\minitwocolumn}
{\footnotesize\baselineskip=12pt
Fig.~1.~
The quark self-energy as the function of the momentum squared
at various dual gluon masses, $m_B=0.3, 0.4$ and $0.5 {\rm GeV}$.
Only a trivial solution is found for the small values $m_B$,
($m_B \stackrel{<}{\scriptstyle \sim} 0.2{\rm GeV}$).
The other parameters are fixed as $e=5.5$ and $a=85{\rm MeV}$.}
\end{minipage}
\vspace{0.3cm}

We study the chiral symmetry restoration at finite temperature using
the imaginary-time formalism.
The finite-temperature SD equation is obtained by making the
following replacement in the SD equation (\ref{sde2}):
{
\setcounter{enumi}{\value{equation}}
\addtocounter{enumi}{1}
\setcounter{equation}{0}
\renewcommand{\theequation}{\theenumi\alph{equation}}
\jot 0.35cm
\barr
p_4 &\rightarrow& \omega_n = (2n+1)\pi T \; ,\\
\int {{d^4 k} \over {(2\pi)}^4} &\rightarrow& T \sum_{m=-\infty}^{\infty}
\int {{{\rm d}{\bf k}} \over {(2\pi)}^3} \; ,\\
M(p^2) &\rightarrow& M(\omega_n ,{\bf p}) \; .
\earr
\setcounter{equation}{\value{enumi}}
}

\noindent
The obtained equation is very hard to solve even
numerically since the quark self-energy depends not only on the three
dimensional momentum $\bf p$,
but also on the Matsubara frequencies $\omega_n$.
We use the {\it covariant-like ansatz} for the quark self-energy at $T\neq0$ as
\beqn
M(\omega_n,{\bf p}) \simeq M_{_T}(\hat p^2)
\eeqn
with ${\hat p}^2 = {\bf p}^2 + {\omega^2_n}$ and $\omega_n=(2n+1)\pi T$.
This ansatz guarantees that the finite-temperature SD equation in the limit
$T \rightarrow 0$ is exactly reduced to the SD equation (\ref{sde2}) at $T=0$.
The final form of the SD equation at $T\neq0$ is derived as
\barr
M_{_T}(\hat p^2)
&=& {T \over {8 \pi^2}} \sum^{\infty}_{m=-\infty}
\int^{\infty}_{\omega^2_m} {d{\hat k}^2} {\int^{1}_{-1}{dz}} {\bf Q}^2
\sqrt{\hat k^2 - \omega^2_m}
{M_{_T}(\hat k^2) \over {\hat k^2+M^2_{_T}(\hat k^2)}} \nonumber \\
&& \times \left[ { 2 \over { {\tilde k^2_{nm}} + m^2_B }  }
+ { {1+\alpha_e} \over {\tilde k^2_{nm}} }
+ {4 \over a}\cdot{1 \over {a+
\sqrt{{\tilde k^2_{nm}}+a^2}}}
\left( { {m^2_B-a^2}
\over {{\tilde k^2_{nm}}+m^2_B} }
+ {a^2 \over {\tilde k^2_{nm}}}
\right) \right] \; ,
\label{sdf}
\earr
where $\tilde k^2_{nm}={\hat k}^2+{\hat p}^2-2z
\sqrt{({\hat p}^2-\omega^2_n)({\hat k}^2-\omega^2_m)}
-2\omega_m\omega_n$.

We solve Eq.(\ref{sdf}) numerically by setting $\omega_{n}=0$ in the
r.h.s. of Eq.(\ref{sdf}), where we use $M(\omega_n,{\bf p}) \approx M(0,{\hat
p})$
being consistent with the {\it covariant-like} ansatz.
We show in Fig.2, the quark self-energy $M_{_T}({\hat p}^2)$ at finite
temperature
rapidly decreases with temperature.
No nontrivial solution is found in the high temperature region,
$T\stackrel{>}{\scriptstyle \sim}110{\rm MeV}$.

\vspace{0.3cm}
\setlength{\minitwocolumn}{4.1in}
\begin{minipage}{\minitwocolumn}
\end{minipage}
\hspace{\columnsep}
\setlength{\minitwocolumn}{2.3in}
\begin{minipage}{\minitwocolumn}
{\footnotesize\baselineskip=12pt
Fig.~2.~
The quark self-energy as the function
of ${\hat p}^2$ at $T=0,~60,~100{\rm MeV}$,
with $e=5.5$, $m_B=0.5{\rm GeV}$ and $a=85{\rm MeV}$.}
\end{minipage}
\vspace{0.3cm}

\noindent
The quark condensate is easily calculated using the quark self-energy
$M_{_T}({\hat k}^2)$ as
\jot 0.5cm
\barr
\langle {\bar q}q \rangle_{_T}
&=& - 4 N_c \cdot T
\sum_{n=-\infty}^{\infty}
\int {{{\rm d}{\bf k}} \over {(2\pi)}^3}
{M(\omega_n ,{\bf k}) \over
{{\bf k}^{2}+\omega_{n}^{2}+M^{2}(\omega_n ,{\bf k})} } \nonumber \\
&\simeq& - {2N_c \over \pi^2}\cdot T
\sum_{n=0}^{n_{\rm max}}
\int^{\Lambda^2}_{\omega_n^2} {d{\hat k}^2} \sqrt{\hat k^2 - \omega^2_n}
{M_{_T}(\hat k^2) \over {\hat k^2+M^2_{_T}(\hat k^2)}}
\earr
with $n_{\rm max} \equiv [ {\Lambda \over {2\pi T}} - {1 \over 2} ]$.
Here, the ultraviolet cutoff is taken to be large enough
as $\Lambda / \Lambda_{\rm QCD} = 10^{3}$.
The quark condensate, $\langle {\bar q}q \rangle
_{_T}$, shown in Fig.3, decreases gradually with temperature at low temperature
and vanishes suddenly near the critical temperature $T_{_C}$.

\vspace{0.3cm}
\setlength{\minitwocolumn}{4.1in}
\begin{minipage}{\minitwocolumn}
\end{minipage}
\hspace{\columnsep}
\setlength{\minitwocolumn}{2.3in}
\begin{minipage}{\minitwocolumn}
{\footnotesize\baselineskip=12pt
Fig.~3.~
The quark condensate at finite temperature
as the function of the temperature divided by the critical temperature
$T_{_C}$.
The same parameters are used as in Fig.2.}
\end{minipage}
\vspace{0.3cm}

\noindent

Finally, we examine the correlation between the critical temperature $T_{_C}$
of the
chiral symmetry restoration and the confinement quantity such as the
string tension $\sigma$. As shown in Fig.4, there exists the strong
correlation between them.

\vspace{0.3cm}
\setlength{\minitwocolumn}{4.1in}
\begin{minipage}{\minitwocolumn}
\end{minipage}
\hspace{\columnsep}
\setlength{\minitwocolumn}{2.3in}
\begin{minipage}{\minitwocolumn}
{\footnotesize\baselineskip=12pt
Fig.~4.~
The critical temperature $T_{_C}$ as the function of the string tension
$\sigma$.
The upper and lower
dotted line are in the cases of the ratio $m_\chi / m_B =2$ and $3$,
respectively.}
\end{minipage}
\vspace{0.3cm}

\noindent
On the physical point of view, higher $T_{_C}$ should be necessary to restore
the
chiral symmetry in the system where the linear-confining potential is much
stronger, because there exists the close relation\cite{{SST},{SST2},{Latt}}
between color confinement and
dynamical chiral-symmetry breaking through QCD-monopole
condensation as was shown in Fig.1.
One estimates as  $90{\rm MeV} \stackrel{<}{\scriptstyle \sim}T_{_C}
\stackrel{<}{\scriptstyle \sim} 130{\rm MeV}$, when one takes the
standard value of the string tension ($\sqrt{\sigma} \simeq 0.45 {\rm
GeV}$).

In summary, we have studied the chiral symmetry using the SD equation in the
DGL theory,
which provides both quark confinement and dynamical chiral-symmetry breaking.
We solve the finite-temperature SD equation numerically with the {\it
covariant-like} ansatz.
The chiral symmetry is restored at $T_{_C}\sim100{\rm MeV}$.
We have found the strong correlation between the critical temperature $T_{_C}$
and the string tension $\sigma$.

\end{document}